\documentstyle[epsf,11pt]{article}

\newcommand{\psc}{\phi_{SC}}
\newcommand{\psd}{\phi_{SD}}
\newcommand{\psun}{\phi_{\odot}(r)}
\newcommand{\psunc}{\phi_\odot(R_\odot)}
\newcommand{\xaxis}{2E\df_{21}}
\newcommand{\yaxis}{2E\df_{31}}
\newcommand{\beq}{\begin{equation}}
\newcommand{\eeq}{\end{equation}}
\newcommand{\bea}{\begin{eqnarray}}
\newcommand{\eea}{\end{eqnarray}}
\newcommand{\barr}{\begin{array}}
\newcommand{\earr}{\end{array}}
\newcommand{\bmat}{\left( \begin{array}}
\newcommand{\emat}{\end{array} \right)}

\newcommand{\bit}{\begin{itemize}}
\newcommand{\eit}{\end{itemize}}

\newcommand{\rttwo}{\sqrt{2}}

\newcommand{\grad}{\nabla}
\newcommand{\B}{{^8}\!B}
\newcommand{\Be}{{^7}\!Be}

\newcommand{\Cl}{{^{37}}\!Cl}

\newcommand{\nres}{(N_e)^{res}}
\newcommand{\ncr}{(N_e)^{cr}}
\newcommand{\nmax}{(N_e)_{max}}
\newcommand{\Psurv}{\bra P(\nu_e \ra \nu_e) \ket}
\newcommand{\bra}{\langle}
\newcommand{\ket}{\rangle}

\newcommand{\nue}{\nu_e}
\newcommand{\nm}{\nu_{\mu}}
\newcommand{\nt}{\nu_{\tau}}
\newcommand{\temu}{{\theta}_{12}}
\newcommand{\tetau}{\theta_{13}}
\newcommand{\tmt}{\theta_{23}}
\newcommand{\th}{\theta}
\newcommand{\thm}{\theta^m}
\newcommand{\ra}{\rightarrow}

\newcommand{\df}{\Delta f}
\newcommand{\dF}{\Delta F}
\newcommand{\dm}{\Delta m^2}

\begin{document}
\title{An Investigation of Equivalence Principle Violations Using
Solar Neutrino Oscillations in a Constant Gravitational Potential}
\author{J.\ R.\ Mureika\thanks{newt@java.usc.edu} ~
\\
{\it Department of Computer Science} \\
{\it University of Southern California} \\
{\it Los Angeles, California 90068~~USA}}

\maketitle

\vskip .25 cm

\textwidth=6in
\noindent
{\footnotesize
{\bf Abstract}

Neutrino oscillations induced by a flavor--dependent violation
of the Einstein Equivalence Principle (VEP) have been recently considered
as a suitable explanation of the solar $\nue$ deficiency.  Unlike
the MSW oscillation mechanism, the VEP mechanism is dependent on
a coupling to the local background gravitational potential $\Phi$.   We
investigate the differences which arise by considering three--flavor
VEP neutrinos oscillating against fixed background potentials, and against
the radially--dependent solar potential.  This can help determine the
sensitivity of the gravitationally--induced oscillations to both 
constancy and size (order of magnitude) of $\Phi$.  In particular, we consider
the potential of the local supercluster, $|\psc|=3\times 10^{-5}$,
in light of recent work suggesting that the varying solar potential
has no effect on the oscillations.  The possibility for arbitrarily large 
background
potentials in different cosmologies is discussed, and the effects of 
one such potential ($\Phi = 10^{-3}$) are considered.
}

\textwidth=7.5in
\section{Introduction}

A resurgence of flavor--oscillation solutions to the Solar Neutrino
Problem has recently inundated the world of particle astrophysics.  
Of these, the two main competitors are
the tried and true MSW mechanism \cite{msw1,wolf}, and the more
radical VEP (for ``Violation of the Equivalence Principle'') 
oscillation model \cite{gasp}.  The MSW mechanism
is often the more seriously considered one, 
due to the basic underlying  hypothesis
of VEP, which requires a generation--dependent violation 
of the Einstein Equivalence Principle ({\it i.e.} flavor--dependent
coupling to the background gravitational potential $\Phi$)
in the neutrino sector, and tends
to give most physicists a headache.  Conversely, there is no
experimental justification for any flavor of neutrino to have
mass, so MSW can also be considered a large assumption.

   Another consideration is the number of flavors with which to work.
Many studies of neutrino oscillations
\cite{bah1,mal1} take the simpler two--neutrino approach by only considering 
the effects of a flavor eigenstate $|\nu_W \ket = (\nue,\nm)^T$.
There are at least three flavors known in nature, though, and
various studies \cite{fogli,me1} have suggested that the two--flavor
model short--changes the viable solution set by considering only
a minute portion of the overall parameter space.  The authors
of \cite{ack} suggest that three--flavors are sufficient to solve
the SNP.

        Proper experimental results from solar neutrino detectors
could help isolate the individual values of the mass eigenvalues
and mixing angles which contribute to oscillations.  Experimental
detection of oscillations induced by the MSW mechanism can yield
determination of the neutrino energy eigenvalue differences
$\dm_{ij}/(2E)$, and hence the neutrino masses $m_i$.  If VEP
is the true underlying mechanism, the analogue of the energy
eigenvalue difference is $2E\Phi\df_{ij}$.  As long as we
are aware of the background potential $\Phi$ that causes the
oscillations, we can isolate the values $\df_{ij}$, and thus
the individual $f_i$s.  While experimental verification of MSW
can teach us new neutrino physics, a detection of VEP oscillations
can additionally yield new information on solar astrophysics and
General Relativity (the universality of the equivalence principle).

       For solar neutrinos, one would generally
expect $\Phi = \phi_{\odot}(r)$, {\it i.e.} that
of the solar interior.  However, there is some debate as to
whether or not the propagating neutrinos feel the effects of
a potential {\em difference}, or whether they feel the overall effects
of the strongest potential \cite{pant1}.  If the latter is the case, then
the gravitational potential of the local supercluster
$|\psc| = 3\times 10^{-5}$ overwhelms the former by almost a
factor of 10.  In effect, the solar neutrino problem has actually
become the {\em solar} neutrino problem: without proper knowledge
of the background potential which contributes to the flavor
oscillations, we can only learn as much new physics
as we could in the MSW case.  That is to say, we can only determine
that neutrinos {\em do} oscillate, but not their gravitational eigenvalues.

	This then raises the question of how much of a difference
the choice of potential makes in the overall process.  The following
 paper will examine the dependence on potential energy
of flavor oscillations in a three--generation framework, from
two specific standpoints.  To begin with, we will
compare the effects of the radially--dependent solar potential
$\phi_{\odot}(r)$ with a constant one of equal magnitude (the solar surface
potential, $|\phi_{\odot}(R_{\odot})| = 2.12\times 10^{-6}$ \cite{weinb}).
This will give an indication of the sensitivity of the data to 
variations over a fixed magnitude.  Secondly, a comparison will be
made of several constant potentials, to examine dependence on varying
magnitudes.  These will primarily include the solar surface potential,
and the local supercluster $|\psc| = 3 \times 10^{-5}$.  An arbitrarily
large potential of the order $\Phi = 10^{-3}$ will be considered,
as well, for reasons discussed in section~\ref{potprob}.

\section{MSW and VEP Oscillation Mechanisms}

	At heart, the MSW and VEP mechanisms are the same.  Both rely on the
existence of two distinct eigenstates of the neutrinos.  In the former case,
there are mass and electroweak eigenstates, $|\nu\ket_M$ and 
$|\nu\ket_W$.  For the latter, the mass eigenstate is replaced with a
gravitational one, $|\nu\ket_G$, as the VEP neutrinos are considered
massless\footnote{There has been some investigation of the possibility
of massive VEP neutrinos, but those are not considered here.  See
\cite{min} for further information.}.  The equations of motion are
diagonal in the mass/gravitational (M,G) basis,

\beq
i\frac{d}{dr}|\nu\ket_{M,G} = H|\nu\ket_{M,G}~,
\label{eqmot}
\eeq
where (\ref{eqmot}) is derivable from the Pauli equation \cite{me1}.
To investigate neutrino flavor dynamics, the (M/G) states
$|\nu\ket_{M,G}$ are rotated to the electroweak basis via
$|\nu\ket_W = V|\nu\ket_{M,G}~,~V \in SU(N)$ for $N$ flavors\footnote{Although
$V$ is a four parameter matrix (3 real rotations and a complex phase), we do
not consider {\tt CP} violations here, and
thus eliminate the complex phase.}.  Due to charged
current interactions in the $\nue$ sector, an additional interaction
${\cal A} = diag(\sqrt{2}G_FN_e,0,0,...)$ must be added, giving the final
equation of motion as

\beq
\i\frac{d}{dr}|\nu\ket_W = \left\{ V H V^{-1} + {\cal A}\right\} |\nu\ket_W =
H' |\nu\ket_W~.
\label{vepeq}
\eeq
for neutrinos of energy $E$.  Here,
${\cal A}(r) = diag(2\sqrt{2}G_FN_e(r),0,0,...)$, with Fermi's
constant $G_F$, and the radially--dependent solar electron density $N_e(r)$.
Subtracting an overall factor of $H'_{11}$ (and hence an unobservable phase
in the wavefunction) from (\ref{vepeq}), and re--diagonalizing,
one obtains eigenstates dependent of either values of $\Delta m_{ij}^2/2E$
or $2E|\Phi|\df_{ij},~ i,j \ne 1, i \ne j$.   The main signature of VEP is
its inverse dependence on the neutrino energy $E$, as compared to MSW.

\section{Three--Flavor VEP Formalism}

A brief review of the underlying formulation of the VEP mechanism
is required before proceeding.  Three--flavor oscillations take place due
to the existence of two non--trivial spinor eigenbases, $|\nu\ket_G = 
(\nu_1,\nu_2,\nu_3)^T~,~ |\nu\ket_W = (\nue,\nm,\nt)^T$, 
related by a rotation matrix $V_3 \in SO(3)$ ({\it i.e.} no {\tt CP} 
violations),

\beq
|\nu\ket_W = V_3 |\nu\ket_G~.
\label{rot}
\eeq

The gravitational eigenbasis evolves according to the massless
Dirac equation

\beq
i D \!\!\!\!/\psi_G = 0~,~\psi_G \equiv \sum_k (|\nu\ket_G)_k \exp\{-iE_kt\}~,
\label{dirac}
\eeq

The Hamiltonian derived from Eq.(\ref{dirac}) becomes off--diagonal
under the change of basis (\ref{rot}),

\beq
H' = V_3 H(\Phi) V_3^{-1} + {\cal A}(r)~,
\eeq
with
\bea
H(\Phi) & = & 2 E |\Phi| \bmat{ccc} f_1 & 0 & 0 \\ 0 & f_2 & 0 \\ 0 & 0 & f_3 \emat~, \nonumber \\
{\cal A}(r) & = & 2 \rttwo G_F N_e(r) diag\{ 1, 0, 0 \}~.
\eea

We explicitly express the dependence on the potential 
of $H = H(\Phi)$ for our later consideration of the various forms 
of $\Phi = \{\psun,\psc,\psd\}$, but omit it
for the other Hamiltonians (and assume its dependence
to be understood in these cases).
The matrix ${\cal A}(r)$ is introduced by the charged--current interactions
of the $\nu_e$s (hence the triviality ${\cal A}_{ij} = 0~~ \forall~ 
\{i,j\} \ne \{1,1\}$), and results in a
shift in the energy eigenvalues $E_k$ from their values in \ref{dirac}.  
A new basis $|\nu\ket_{MG}$  and set of corrected eigenvalues
$\{E''_1,E''_2,E''_3\}$ may be found by rediagonalization

\beq
H'' = V_3^m H' (V_3^m)^{-1}~,
\eeq

where $V_3^m=V^m_3(\temu^m,\tetau^m,\tmt^m)$ is the corrected rotation 
matrix, with matter--enhanced angles (see \cite{me1} for the definition
of these, or \cite{zag,barg} for their MSW equivalent).

The wave equation may now be solved, subject to the modified Hamiltonian
$H''$, and the subsequent survival probabilities for neutrino eigenstates
can be found.  For solar neutrinos, the value of interest is $\Psurv$,
since most detectors to date are sensitive only to first--generation
neutrinos (this is not true of future detectors such as Superkamiokande
or SNO, which can detect neutral current interactions from all flavors).
Averaged over the Earth--Sun distance, the expression obtained is

\bea
\Psurv& =& \sum_{i,j=1}^{3} |(V_3)_{1i}|^2\; |(P_{LZ})_{ij}|^2\; |(V_3^m)_{1j}|^2 \nonumber \\
 & = & c_{m12}^2 c_{m13}^2 \left\{ (1-P_1)c_{12}^2
 c_{13}^2 + P_1 s_{12}^2 c_{13}^2 \right\} \nonumber \\
& & + s_{m12}^2 c_{m13}^2 \left\{ P_1 (1-P_2) c_{12}^2 c_{13}^2 + (1-P_1)(1-P_2)
s_{12}^2 c_{13}^2 + P_2 s_{13}^2 \right\} \nonumber \\
& & + s_{m13}^2 \left\{ P_1 P_2 c_{12}^2 c_{13}^2 + P_2 (1-P_1) s_{12}^2
s_{13}^2 +(1-P_2) s_{13}^2 \right\}~. \nonumber \\
{  }
\label{p3}
\eea

which is a function of the matter--enhanced mixing angles $\thm_{13},
\thm_{13}$ and eigenvalues differences $\dF_{21}, \dF_{31}$, and 
the regular (vacuum) equivalents $\th_{12}, \th_{13}$, and 
$\df_{21}, \df_{31}$.  Note that $\Psurv$ does not depend on
interactions between the second and third flavors, $\nm \ra \nt,
\nt \ra \nm$.

The results of \cite{me1,me2} offer insights into the fundamental
theoretical and observational differences between the VEP and MSW
mechanisms.  The overall conclusion was found to be that the
addition of the third neutrino flavor can significantly enhance
the overall allowed parameter space from that of the two--flavor
model.  Unique experimental signatures in observed $\nue$ fluxes
can also arise between the VEP and MSW, which could be used to
differentiate between mechanisms.

\section{A Potential Problem!}
\label{potprob}

While most papers on VEP tend to agree on the form of the oscillation
mechanism\footnote{There is a brief mention in \cite{bah1} of an 
alternative form of equivalence principle violation, based on
neutrinos coupling to $\grad\Phi$, instead of $\Phi$, in some
string theories.}, there is some debate as to which background potential
is at work.  Most works have assumed that neutrinos feel the
potential difference of the sun from their creation points, to
their exit on the surface, which ranges between $\; \sim (-7, -2) 
\times 10^{-6}$ \cite{me1}.
Alternatively, \cite{pant1} suggests that the neutrinos feel 
instead the constant background potential of the Local Supercluster 
(Great Attractor), which in the neighborhood of the solar system is 
$|\psc| = 3\times 10^{-5}$.  This overpowers the surface potential 
of the Sun by almost a factor of 10.  Results from the COBE data
suggest that the dominant local background potential is very close
to this value \cite{robb3}.

Furthermore, while the current dominant potential in the
local neighborhood seems to be that of $\psc$, one should be
careful to consider the overall potential contribution of 
the Universe.  Following \cite{good}, the local background
potential can be expressed as the sum of all contributing
potentials, $\Phi_{local} = \phi_{\oplus} + \phi_{odot} + \phi_{MW} + \psc
+ \phi_U + C$.  The contributors of the potentials are, respectively,
the Earth, Sun, Milky Way, Local Supercluster, the Universe, and
an arbitrary constant $C$.  This is different from Equation~(3) of \cite{good}
in that the (dominant) potential $\psc$ has been added.  The
constant $C$ comes from an arbitrary cosmological model that
might be at work, but is undetectable by our standards.
That is, as expressed in \cite{good}, ``we cannot `step outside
the universe' to [measure the actual value of $C$]''\footnote{This
can be thought of as a type of philosophical argument, along
the lines of ``how do we know the reality is not just an illusion?'',
but with actual physical evidence to support it!}
Hence, a future cosmology might contain a $C$ such that $\Phi_{local}
\sim C$.  If $C$ is sufficiently large, it will overwhelm the
other contributors.  For this work, we will assume the value
of $C = 10^{-3}$, and denote this by $\psd$.

  From solar neutrinos alone, we can only hope to isolate the
values of $\df_{ij}\Phi$ (see, for example, \cite{robb}, 
and not the individual eigenvalues
$\df_{ij}$.  According to the aforementioned product, the values
of $\Phi$ and of $\df_{ij}$ are inversely linked.  So, by
considering the effects of the Local Supercluster $\psc$ over
the effects of the solar potential $\psun$, one is effectively
shifting the values of the violation parameters down by an order
of magnitude.  At first glance, this would seem to not be a big
problem, but it should be noted that shifts of this magnitude
can strongly affect the oscillation behavior for critical
values of mixing angle and EEP--violations from the other flavor.

\section{$\Psurv$ Curves and Surfaces}

	For two--flavor analyses of neutrino oscillations, it is 
useful to study the averaged $\nue$ survival probability curves
$\Psurv$.  Such plots are generally expressed as functions of
the argument $E\df$, scaled to an appropriate value.  
These curves indicate the overall behavior of
$\Psurv$ with respect to values of violation parameters
$\df$ ({\it i.e.} $f_1$ and $f_2$) and mixing angle $\th$. 
In particular, it is of use to
examine the small and large--mixing angle limits.

	For two flavors, the resonant boundary for solar neutrinos is 
determined by equating the resonance density
\beq
(N_e)_{res} = \frac{\sqrt{2} |\psun|E\df\cos{2\th_G}}{G_F}~,
\label{res}
\eeq
Since the density of the sun is strictly decreasing from its center,
then if the resonance density for a neutrino is greater than the density
at which it was created, it will never undergo matter--enhanced oscillations
and will propagate as in vacuum (note that they may still undergo vacuum 
oscillations).  Using the radial solar potential
$\psun$, one obtains a resonance boundary of $\df \sim 2 \times 10^{-12}$ for
neutrinos of energies $E_{\nu} > 0.2\;$MeV \cite{bah1}.

	An extension to a three flavor model requires something
extra, though.  Instead of just one specific curve for $\temu$
(the equivalent of the two--flavor angle $\th$), there can be
a whole family of curves $\{\Psurv\}$ corresponding to the
same $\temu$.   These are, of course, determined by different
values of $\tetau$, and imply that the large and small limits
of $\tetau$ must also be considered.   Hence, it becomes apparent
that an examination of a {\em probability surface} is in order.

	These surfaces are of use in three--flavor oscillation
dynamics as the demonstrate the full range of behavior of $\Psurv$
for all values of $\df_{i1}\Phi$ for a given set of $\temu,\tetau$.
This includes regions for which neutrinos  with resonance densities
above the maximal solar electron density $(N_e)^i_{max}=$100$g\,cm^{-3}$.
Again, if the resonance density for a neutrino $i$ exceeds this value, the
neutrino will not feel matter--enhancement, and will propagate as
in vacuum.  The resonance density for each flavor is determined by 
the generalization of (\ref{res}) to three flavors,
by replacing $\df$ with $\df_{ij}$ for the
$\nu_i \ra \nu_j$ resonances. 

	Note that the region encompassed by these surfaces includes
areas where $\df_{21} > \df_{31}$.  This corresponds to a broken
hierarchy in the eigenstates ({\it i.e.} $f_2 > f_3$).  We would like
to think that nature would follow a hierarchial ladder between 
generations, but there is nothing which physically rules out the
possibility of breaking this.  Indeed, the behavior of $\Psurv$
does exhibit noticeable changes when the hierarchy is broken.
In particular, this generates a change in the values of
the matter--enhanced mixing angles $\thm_{21}$ and $\thm_{31}$.
This change is less apparent for small vacuum mixing angles,
while it can create sizable differences for large vacuum 
angles \cite{me1}.

	It should be
noted that these plots show the survival probabilities for
$\B$ $\nue$s\footnote{These originate from the thermonuclear decay
reaction $\B \ra \Be + \nue$; for more information on solar
neutrino--producing reactions, see \cite{bah2}.}, the easiest
of the solar neutrino spectrum to detect.  The majority of the
events recorded in the Homestake ($\Cl$), as well as all events
detected by Kamiokande, are of this type, due to their extremely
high ranging energy spectrum of $[0,15]\;$MeV.
The figures herein have been calculated using the most recent
solar neutrino data and Standard Solar Model data available
\cite{bah4,bah5}.

	The expression for $\nres$ in (\ref{res}) is clearly a function
of the background potential $\psun$.  Rearranging this, it is obvious 
that 

\beq
\df = \frac{(N_e)_{max} G_F}{\sqrt{2} E |\psun|}~,
\eeq
{\it i.e.} the resonance boundary varies inversely with the potential.
So, there should be a decreasing limits on the values of $\df$ which permit
resonance as the potential increases.  For the cases we concern ourselves
with here, these boundaries should approximately (within an order of magnitude)
occur at $\df \sim 10^{-11}$ for $\psc$ and $\df \sim 10^{-13}$ for $\psd$.  

	In addition to the
$\B$ $\nue$s, the other main type of detectable neutrino is
the low energy pp-neutrino, resulting from the proton--proton
chain reaction.  These are emitted over an energy spectrum 
ranging to $0.42$MeV.  Since their maximal energy is at least
one order of magnitude different from the maximal $\B$ neutrino
energy, then the overall product $E\df_{ij}$ can shift by
the same order of magnitude.  It was observed in \cite{me1}
that this shift can result in different resonant behavior
between the reaction types of solar neutrinos, such that
some can exist under the resonant barrier 
($2E\df_{ij}/10^{-18}$MeV$\approx 10^{6}$), while others can be
over.  The crux of this is that certain neutrino
types will be subject to resonant behavior (for $2E\df_{i1} <
10^{6}$, in units of $10^{18}$MeV$^{-1}$), while certain 
types will propagate as in vacuum ($2E\df_{i1} > 10^{6}$).
While this analysis is restricted to $\B$ neutrinos, the 
interested reader should reference \cite{me1} for an appropriate
discussion.

\section{Discussion of Figures}

	As previously mentioned, this work considers two distinct
cases of survival probabilities, as a function of background
potential energy.  We first consider the dependence on $\Psurv$
curves as a function of variability in potential over a 
specific order of magnitude, and secondly as a function of
variability of order of magnitude.

\subsection{$\psun$ {\it v.s.} $\psunc$}
\label{var}

\begin{figure}
\leavevmode
\epsfysize=470pt
\epsfbox[85 70 530 690] {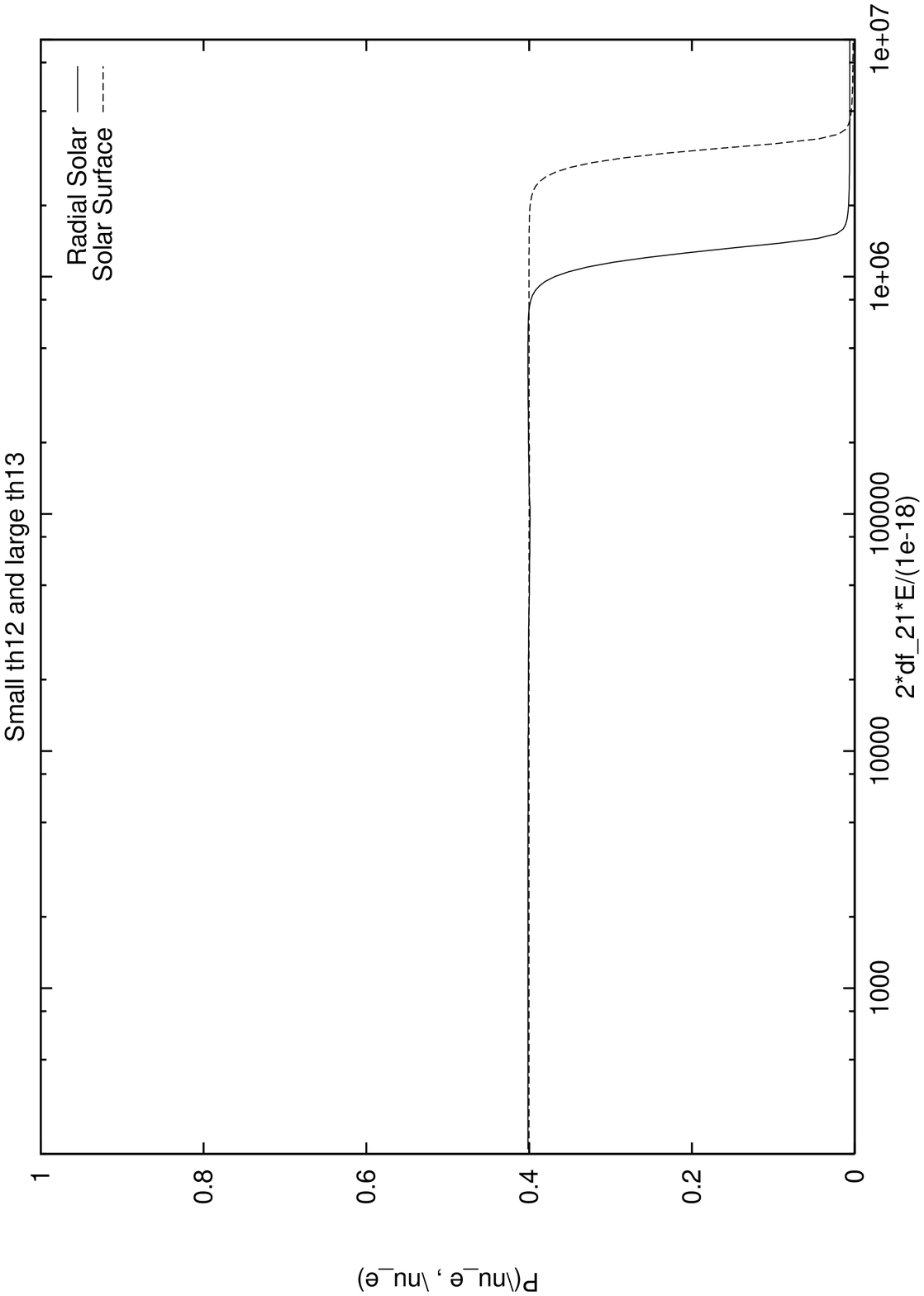}
\caption{}
\label{sunssl}
\end{figure}

\begin{figure}
\leavevmode
\epsfysize=470pt
\epsfbox[85 70 530 690] {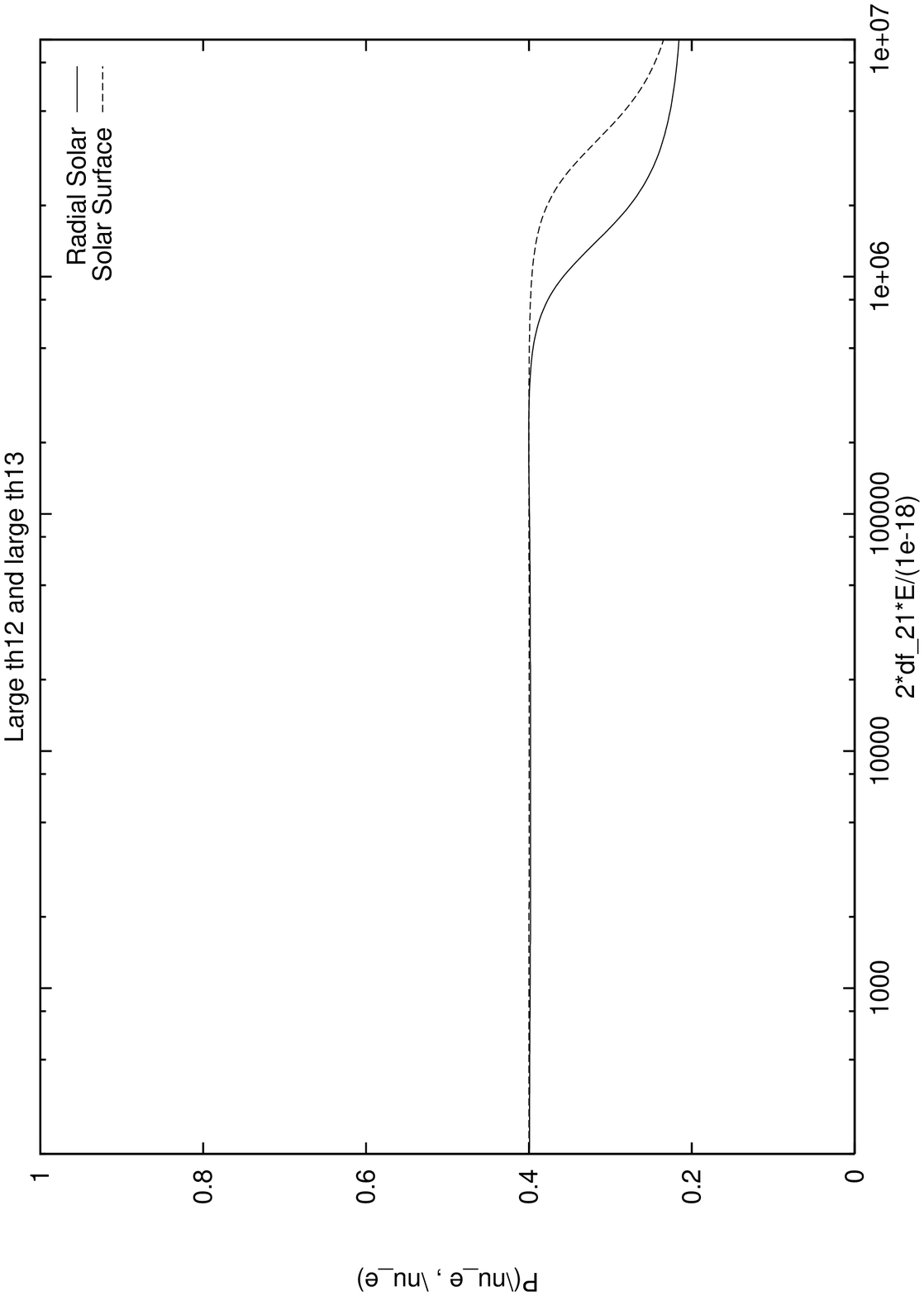}
\caption{}
\label{sunsll}
\end{figure}

	It was noted in \cite{me1} that there is a strong 
influence from the $\tetau$ contribution of the third flavor.
In particular, for large $\tetau$, the survival probabilities
for $\nue$ become effectively $\Psurv \sim \sin^2{\tetau}$.
This is clearly evident from Figures~\ref{sunssl},\ref{sunsll},
which demonstrate the behavior of $\Psurv$ from solar neutrino
data in the small and large $\temu$ range, for small $\tetau$.
The end result of the difference in potential presents itself
as a horizontal shift to the right for $\Psurv(\psunc)$ when
the slices are considered.  The shape of the curve is preserved, 
which suggests that the
change in potential has merely shifted the resonance density
by an overall constant.  Both curves assume the same values
before and after this barrier.  The portions of the curve
for which $2E\df_{21} > 5\times 10^{6}$ correspond to
broken hierarchy between the violation parameters $\df_{i1},
i=2,3$. 

\begin{figure}
\leavevmode
\epsfysize=470pt
\epsfbox[85 70 530 690] {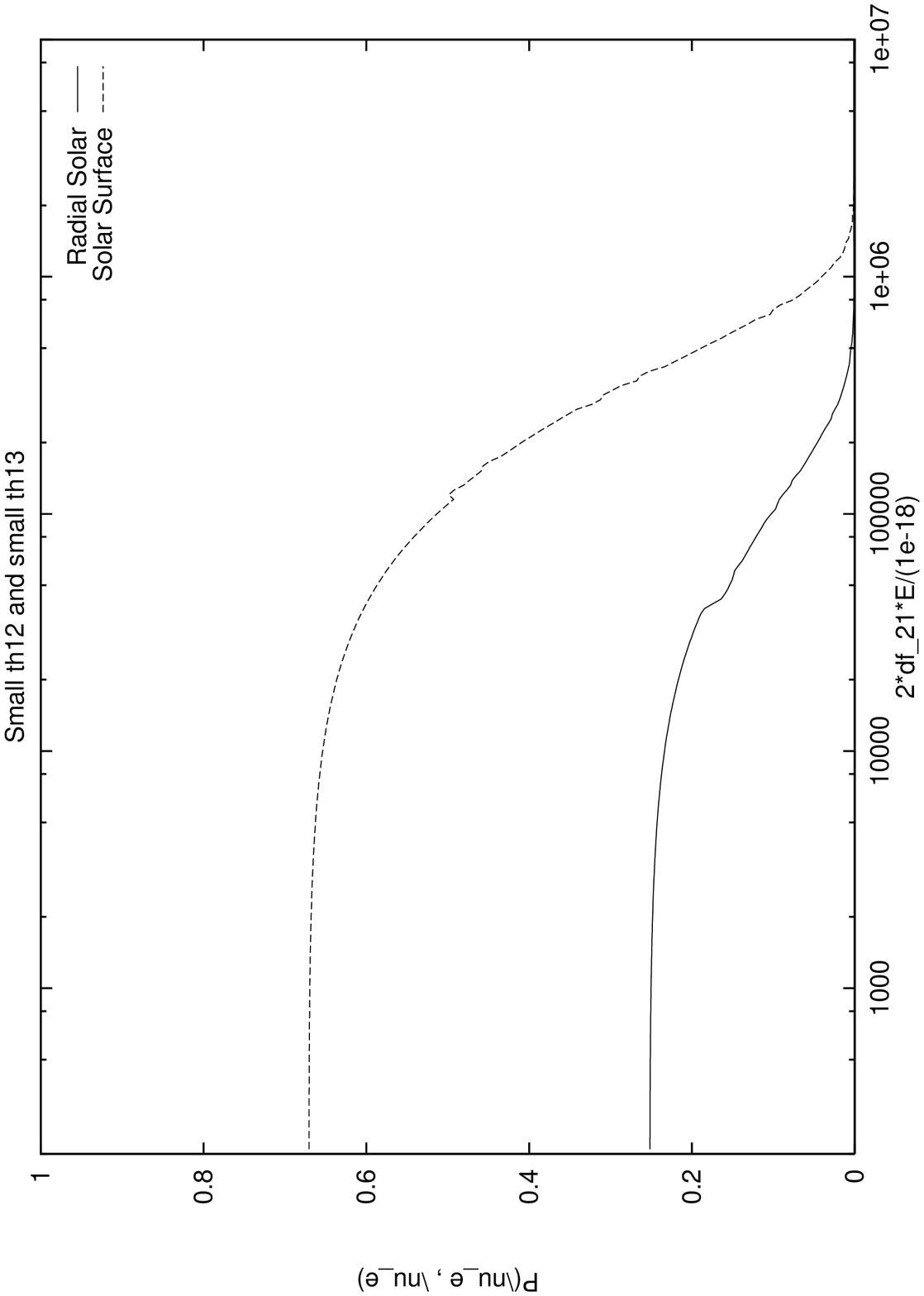}
\caption{}
\label{sunsss}
\end{figure}

\begin{figure}
\leavevmode
\epsfysize=470pt
\epsfbox[85 70 530 690] {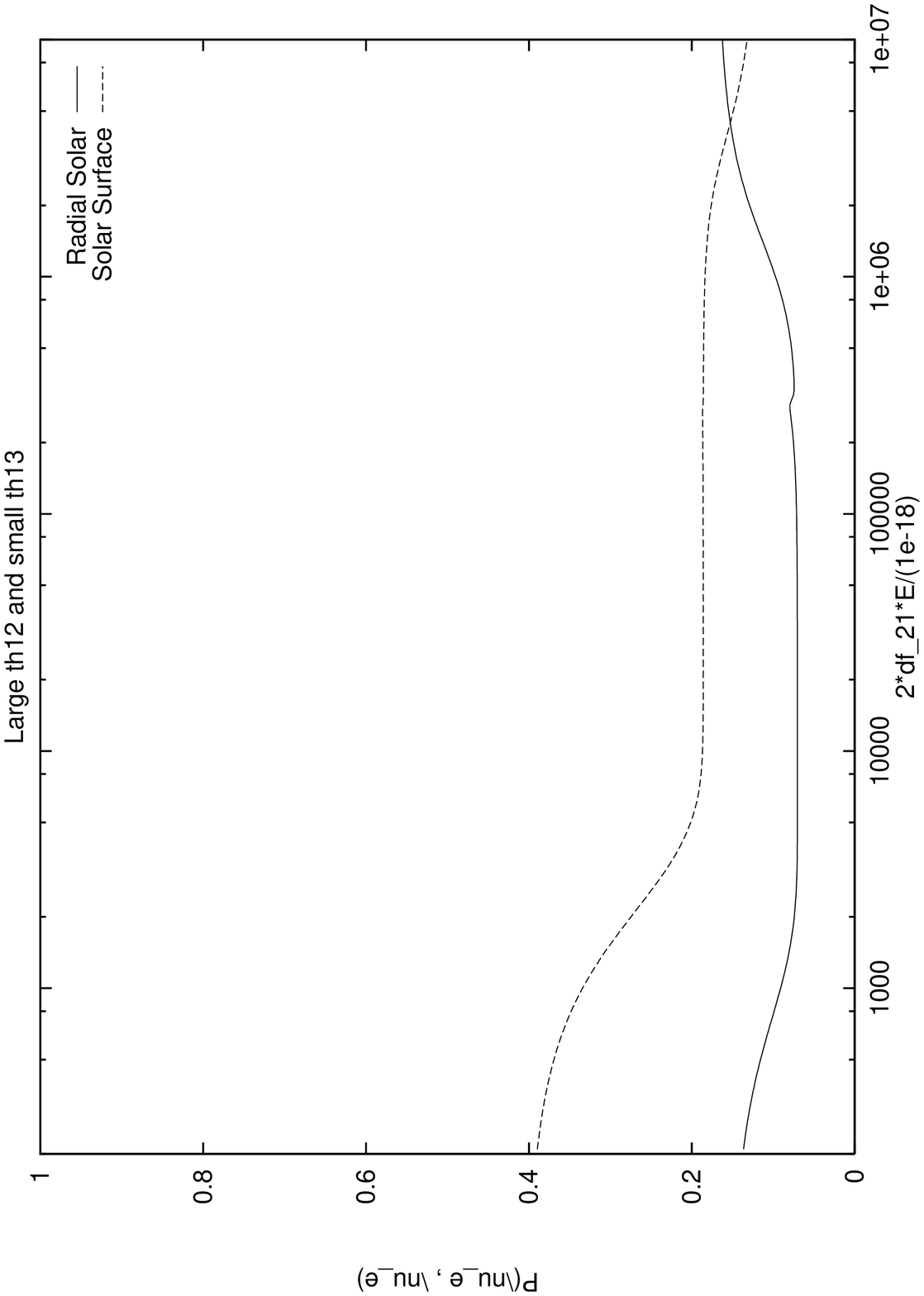}
\caption{}
\label{sunsls}
\end{figure}

	Meanwhile, the story for small $\tetau$ is set to
a somewhat different tune.  We observe vastly different behavior
for both curves in each case of small and large $\temu$ 
(Figures~\ref{sunsss},\ref{sunsls}).  Both plots show that the
the survival probability $\Psurv(\psunc)$ is greater than
$\Psurv(\psun)$.  The shift in the location of the resonance
boundary 

\subsection{Order of Magnitude Dependence}
\label{mag}

	In contrast to the previous section, there is a noticeable
difference in $\Psurv$ curves and surfaces when the order of magnitude
of the effective potential $Phi$ is considered.  This shift has
large implications in particular when it comes to deciding which
potential is at work, if the VEP mechanism is correct.  Indeed,
the evidence presented here could help point to the correct choice,
subject to experimental verification ({\it e.g.} $\B$ flux spectra
can be directly computed with knowledge of $Psurv$).

\begin{figure}
\leavevmode
\epsfysize=470pt
\epsfbox[85 70 530 690] {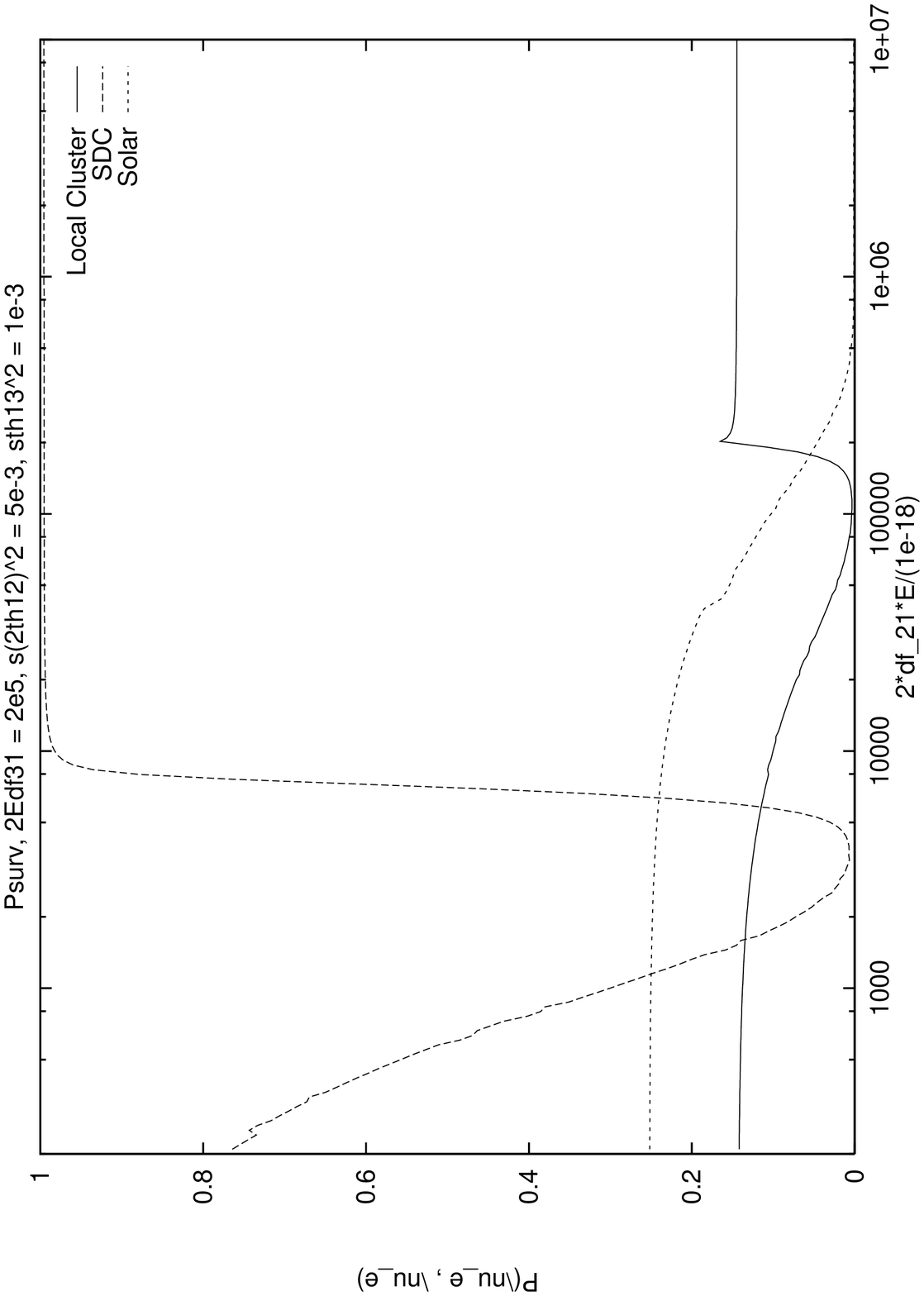}
\caption{}
\label{3fsmsm}
\end{figure}

\begin{figure}
\leavevmode
\epsfysize=470pt
\epsfbox[85 70 530 690] {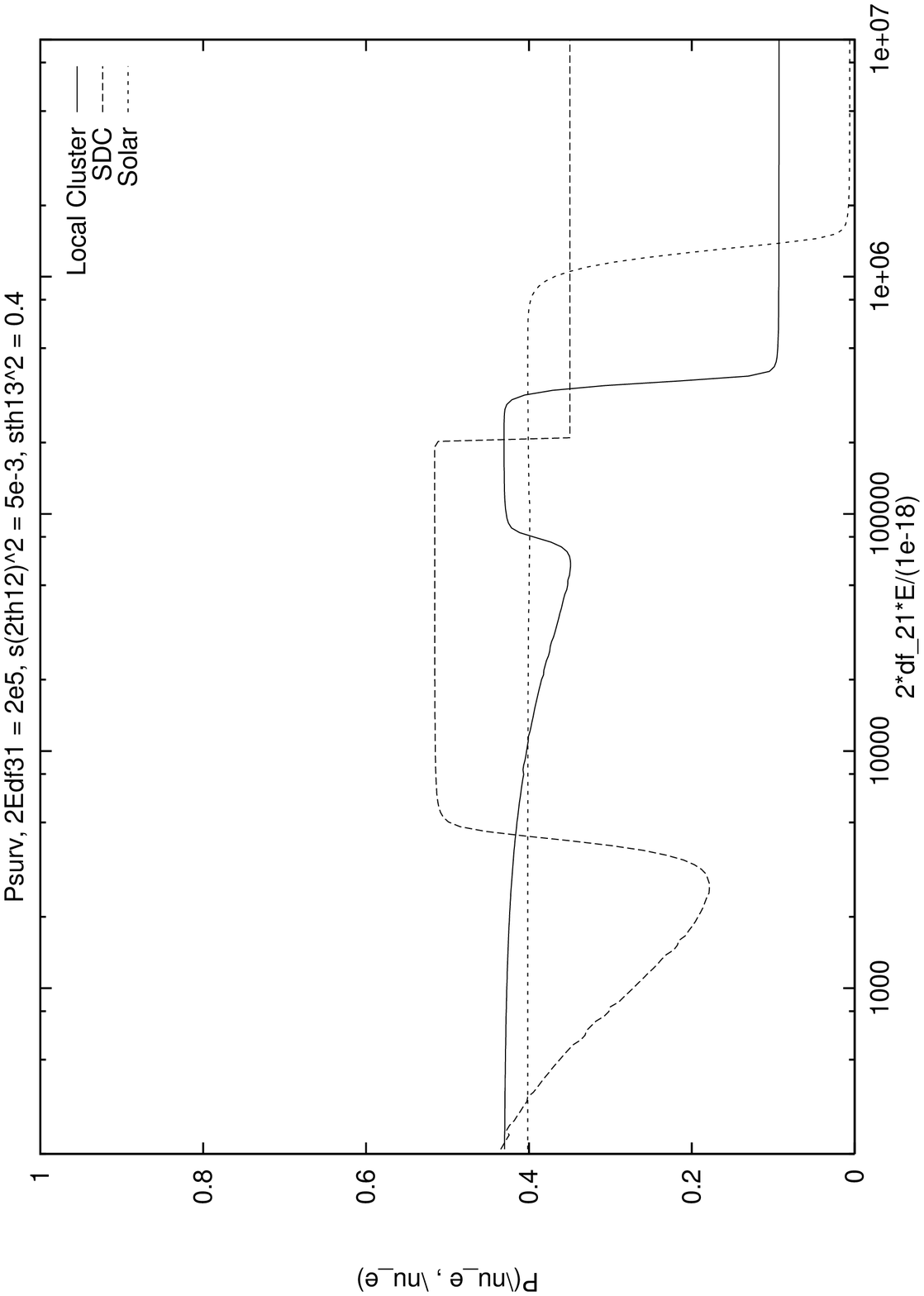}
\caption{}
\label{3fsmlrg}
\end{figure}

	For the purposes of this discussion, both the survival
probability curves and surfaces will be examined for select 
choices of the potential.  As in Section~\ref{var}, the small
and large $\temu$ (fixed $\tetau$ large)  curves are presented
in Figures~\ref{3fsmsm},\ref{3fsmlrg}.  The chosen potentials here
are $\psun,\psc$, and $\psd$.  It is assumed that $\psunc$ will
vary as discussed in the previous section, and thus is not considered
here.

\begin{figure}
\leavevmode
\epsfysize=470pt
\epsfbox[85 70 530 690] {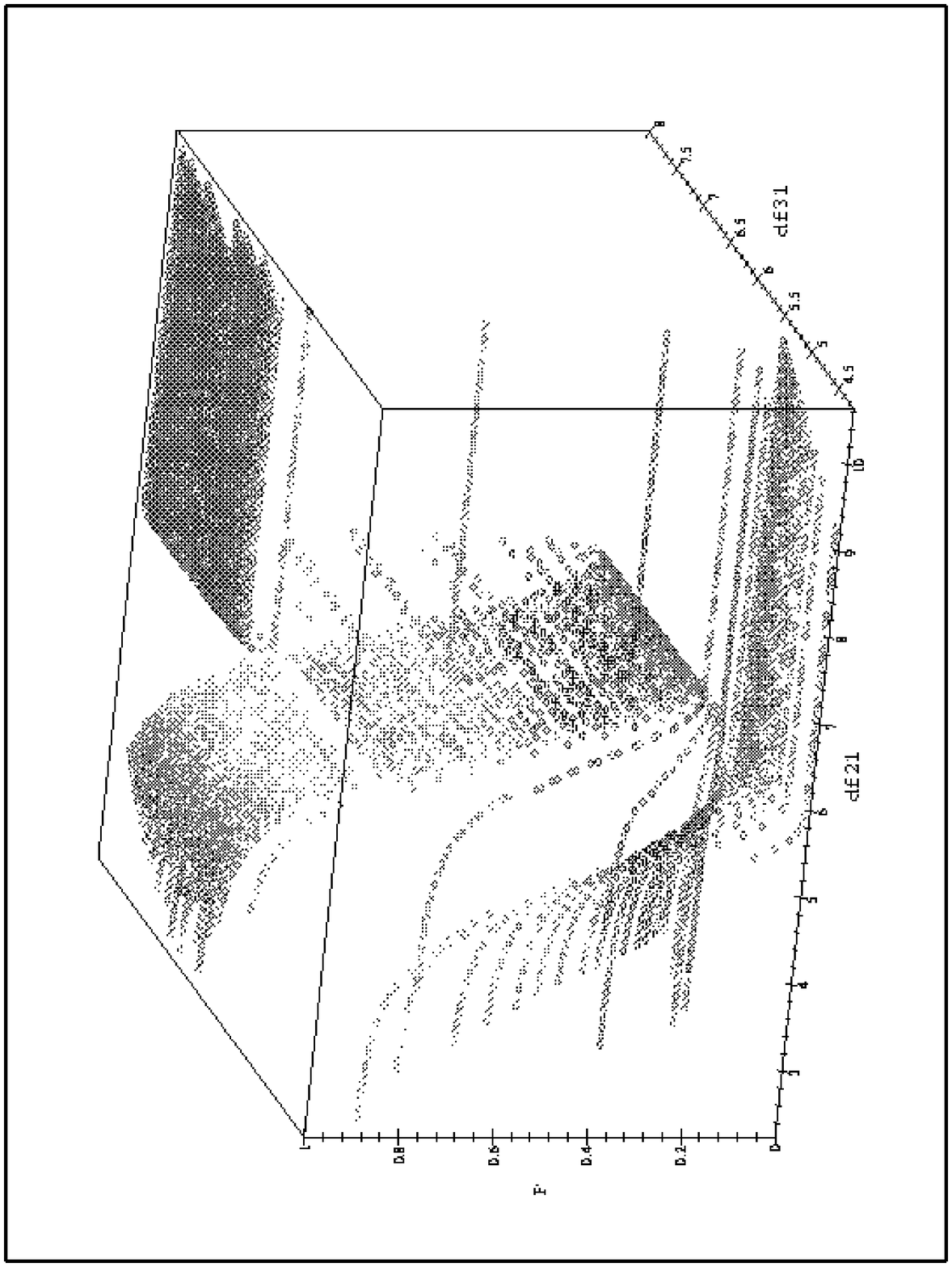}
\label{3dsssun1}
\caption{}
\end{figure}

	Figure~\ref{3fsmsm} represents the small--angle mixings
for both 12-- and 13--resonances.  These curves correspond to a choice 
of $\df_{31} = 10^{-13}$, at a fixed value of $\yaxis = 2\times 10^{6}$.
In this case, each potential induces a drastically different shape
for each curve.  The choice of $\yaxis$ places the Solar curve in
the double--resonance region (the $\nue$ can resonate to both
$\nm$ and $\nt$), while the $\psd$ curve allows only $\nue \ra \nm$
resonances.  In fact, the shape of this curve approaches that for
two--flavor non--adiabatic resonance \cite{me1}.  By referencing
Figure~\ref{3dsssun1}, it is evident that the plots for each 
$\Phi$ represent equivalent slices for $\psun$ at different
values of $\yaxis$ (in this case, roughly in the range $5-7$).

\begin{figure}
\leavevmode
\epsfysize=470pt
\epsfbox[85 70 530 690] {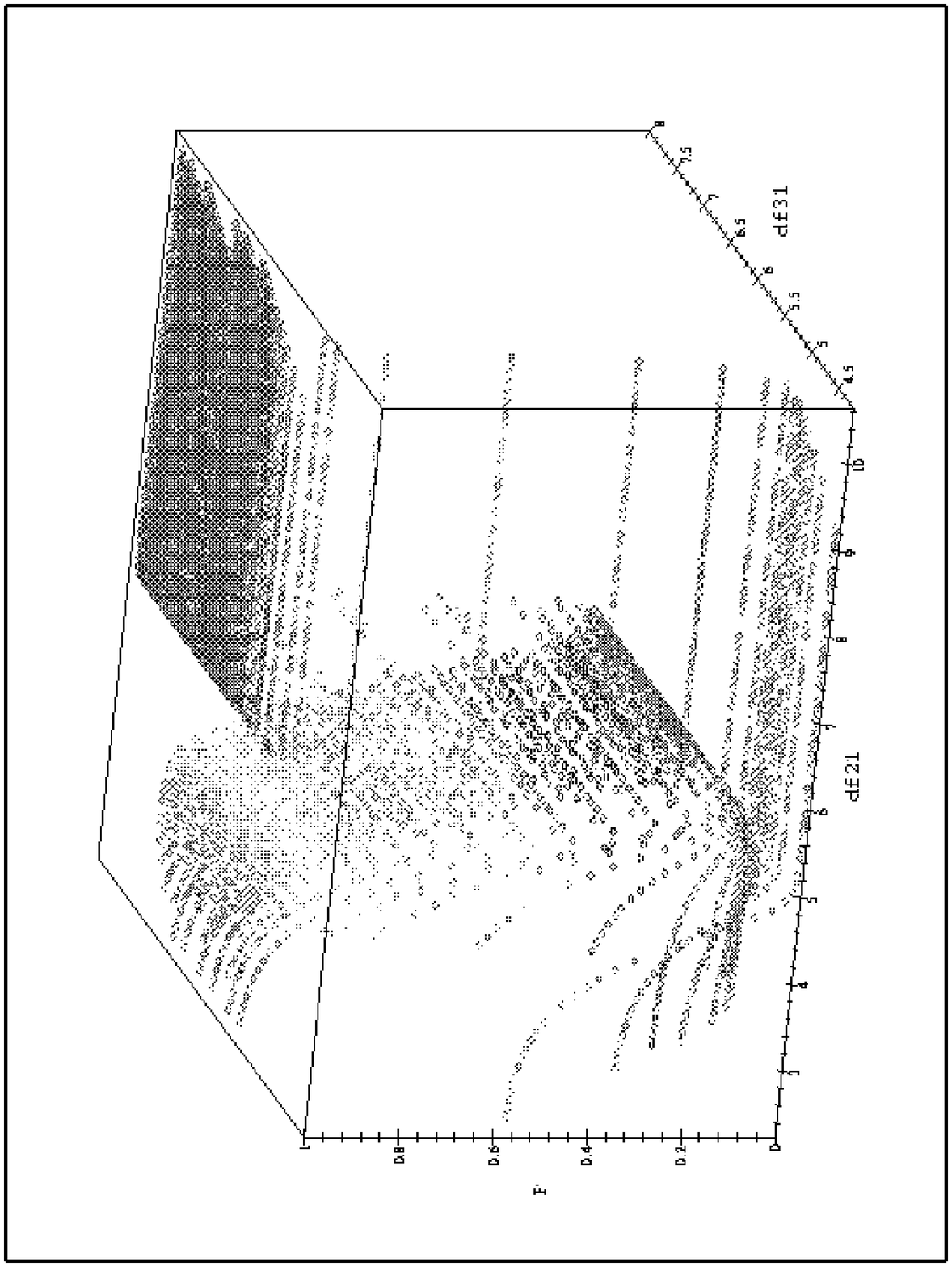}
\caption{}
\label{3dssclu1}
\end{figure}

\begin{figure}
\leavevmode
\epsfysize=470pt
\epsfbox[85 70 530 690] {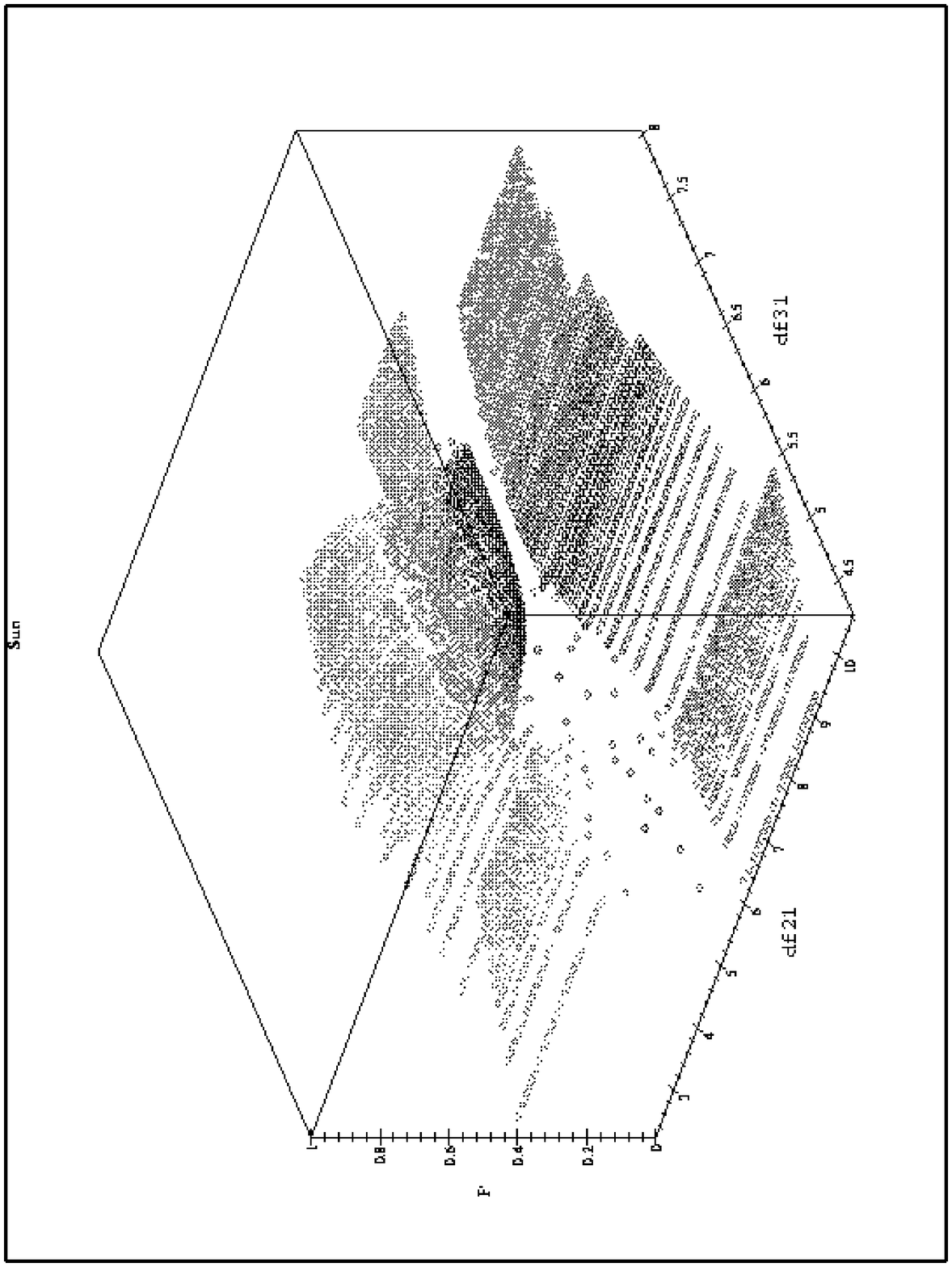}
\caption{}
\label{3dslsun1}
\end{figure}

\begin{figure}
\leavevmode
\epsfysize=470pt
\epsfbox[85 70 530 690] {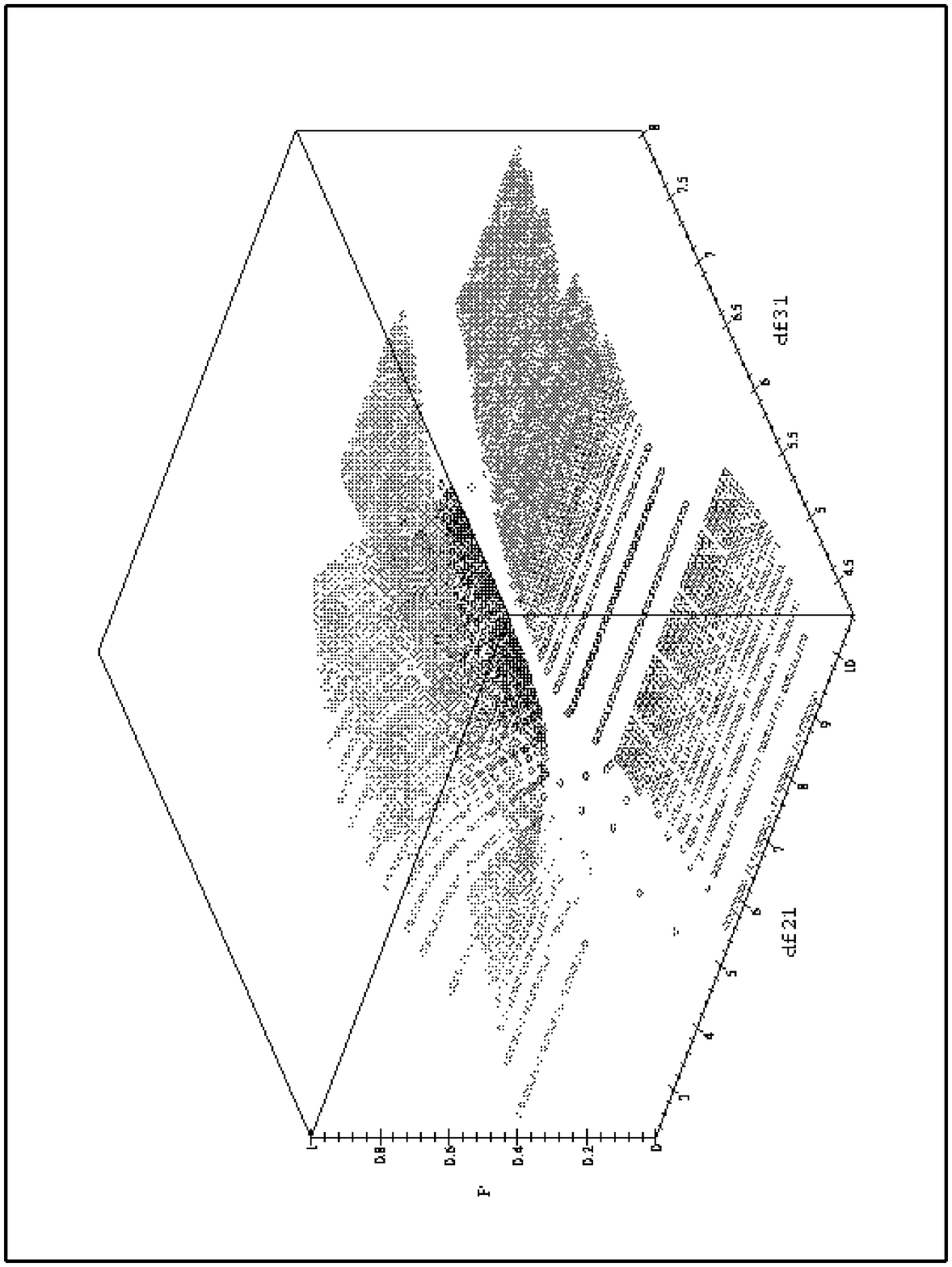}
\caption{}
\label{3dslclu1}
\end{figure}

	A similar analysis hold true for the other cases of small
and large $\temu, \tetau$, and thus will not be discussed further.
To support the claim that the $\Psurv$ surfaces for each value
of $\Phi$ are shifting in parameter space, Figure~\ref{3dssclu1}
is offered for comparison with Figure~\ref{3dsssun1}.  This
is the resulting surface for $\Phi = \psc$.  In addition, 
Figures~\ref{3dslsun1},\ref{3dslclu1} present the case for $s_{12}^2 = 0.8,
s_{13}^2 = 0.001$.  It is clear from both comparisons that the
overall shapes of the surfaces are the same, up to a shift of an
overall constant.  As suggested, this corresponds to an order
of magnitude shift in the resonance boundary.  

\begin{figure}
\leavevmode
\epsfysize=470pt
\epsfbox[85 70 530 690] {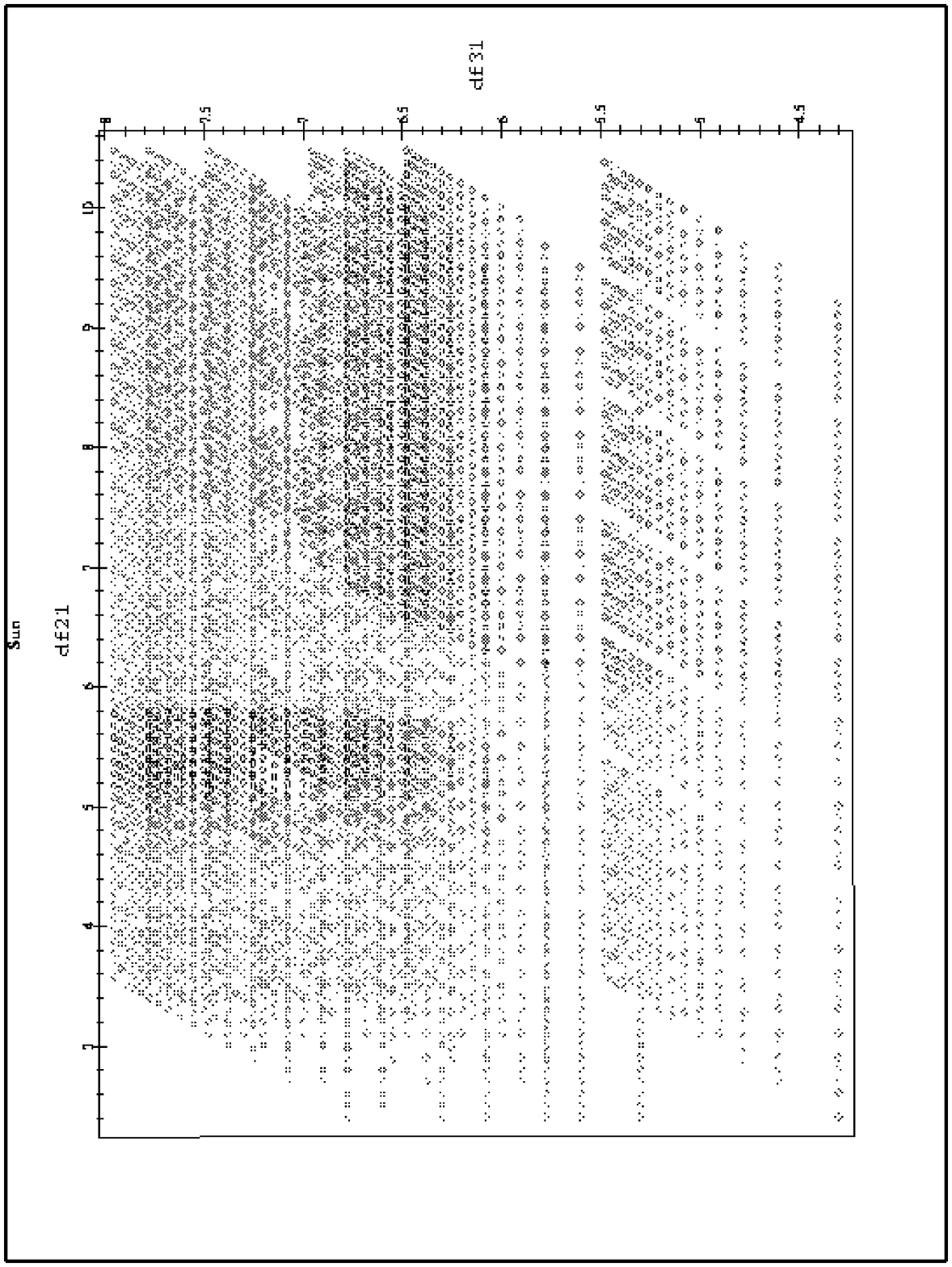}
\caption{}
\label{sunoh1}
\end{figure}

\begin{figure}
\leavevmode
\epsfysize=470pt
\epsfbox[85 70 530 690] {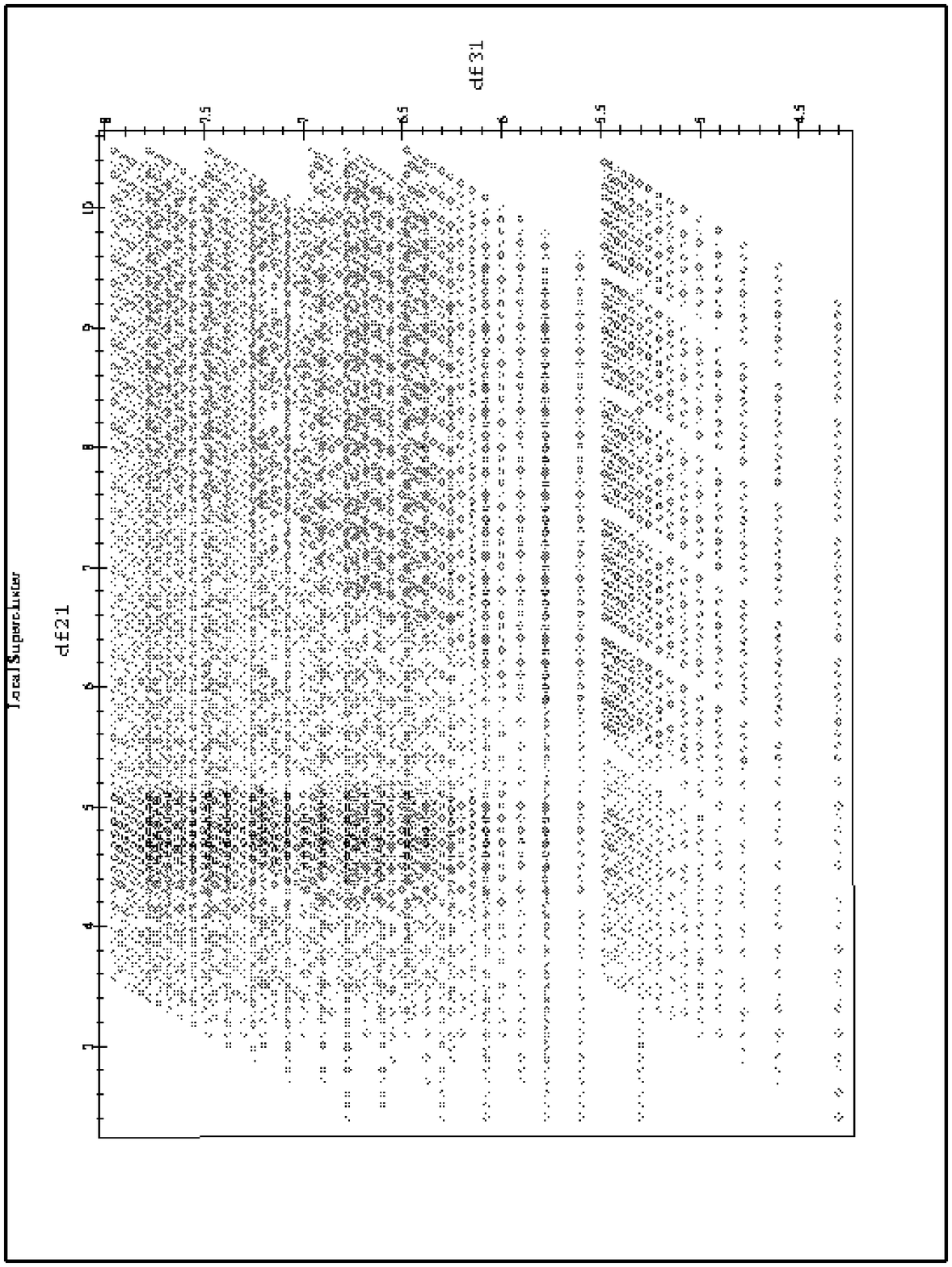}
\caption{}
\label{cluoh1}
\end{figure}

Figures~\ref{sunoh1} and \ref{cluoh1} are overhead projections of
Figures~\ref{3dslsun1},\ref{3dslclu1} in the ($\xaxis$, $\yaxis$) plane.
The lighter--shade triangular--shaped wedge at the center depicts two of the
features discussed here.  The vertical boundary represents the
resonance boundary, where $\ncr > \nmax$ for each neutrino,
while the diagonal one shows the intersection of the plane
$\df_{21} = \df_{31}$.  It can be seen that the former shifts as discussed
according to the potential $\Phi$, while obviously the latter
remains fixed. 

\section{What's All This, Then?}

	It thus becomes apparent that the same type of shifting effect
as that noted in \cite{me1} could conceivably occur, depending on the 
choice of gravitational potentials affecting the VEP oscillations.  
Whereas before it was thought that VEP oscillations were
singly dependent on the product $\Phi\df_{ij}$, these results
suggest that perhaps the value of the background potential
$\Phi$ can indeed be constrained by the resultant neutrino
data.  The figures presented here immediately determine the
observed energy spectrum of $\B$ neutrinos, and many future
neutrino observatories could detect these variations.  As with
the results in \cite{me1,me2}, it is found that the shapes of
the probabilities are strongly determined by the size of $\tetau$,
and small 13--angle oscillations provide a much more diverse
spectrum of $\Psurv$ curves for different input values (in this
case, the potential $\Phi$).

	Alternatively, what could also solve such a problem,
should VEP be the mechanism at work, is
the detection and subsequent spectrum analysis of extra--solar
neutrinos from sources whose gravitational potential is known
to either great accuracy or reliability. Various papers have addressed 
the detection of 
such high--energy intergalactic neutrinos subject to the VEP
mechanism \cite{min3}.  

	Additionally, the discovery of a possible ``gravitationally--
induced quantum phase'' was recently discussed in the literature
\cite{phase1}.  While the reference treatment deals with MSW
neutrinos which can possibly experience a phase shift due to
interactions with strong gravitational sources, it may be possible
to extend the analogy to VEP neutrinos \cite{me3}.  Should such
an extension be possible, then it may be possible to determine
the potential felt in the vicinity of the Sun, since the gravitationally--
induced quantum mechanical phases of \cite{phase1} would be functions
of the source potential (hence the product $\df_{ij}\Phi_{Source}$,
and not just the product $\df_{ij}\Phi_{Solar System}$.  These
conclusions are as of yet unverified.

\vskip .25 cm

\noindent
{\bf Acknowledgements}
 
I thank R.\ B.\ Mann profusely for proof--reading the manuscript,
for offering great insight and suggestions, and for allowing me to
perform calculations on his computer.

\pagebreak
\noindent
{\bf Figure Captions} 

\noindent The following conventions are used for the figure conventions.
The notation {\tt df21, df31} on the plots corresponds to $\xaxis$,$\yaxis$
which are expressed in units of $10^{18}$~MeV$^{-1}$.  Also, 
$P = \Psurv$, and $s_{2\temu}^2 = \sin^22\temu$, $s_{\tetau}^2 =
\sin^2\tetau$.\\

\noindent{\bf Figure 1:} $\Psurv$ curves for $s_{2\temu}^2 = 5\times 10^{-3}, 
s_{\tetau}^2 = 0.4$, $\yaxis = 2\times 10^{6}$, $\Phi = \psun, \psunc$.\\

\noindent{\bf Figure 2:} $\Psurv$ curves for $s_{2\temu}^2 = 0.8,
s_{\tetau}^2 = 0.4$, $\yaxis = 2\times 10^{6}$, $\Phi = \psun, \psunc$.\\

\noindent{\bf Figure 3:} $\Psurv$ curves for $s_{2\temu}^2 =  5\times 10^{-3},`
s_{\tetau}^2 = 10^{-3}$, $\yaxis = 2\times 10^{6}$, $\Phi = \psun, \psunc$.\\

\noindent{\bf Figure 4:} $\Psurv$ curves for $s_{2\temu}^2 = 0.8,
s_{\tetau}^2 = 10^{-3}$, $\yaxis = 2\times 10^{6}$, $\Phi = \psun, \psunc$.\\

\noindent{\bf Figure 5:} $\Psurv$ curves for $s_{2\temu}^2 = 0.8,
s_{\tetau}^2 = 10^{-3}$, $\yaxis = 2\times 10^{6}$, $\Phi = \psun, \psunc,
\psc, \psd$.\\

\noindent{\bf Figure 6:} $\Psurv$ curves for $s_{2\temu}^2 = 0.8,
s_{\tetau}^2 = 10^{-3}$, $\yaxis = 2\times 10^{6}$, $\Phi = \psun, \psunc,
\psc, \psd$.\\

\noindent{\bf Figure 7:} $\Psurv$ surface for $s_{2\temu}^2 = 5\times 10^{-3},  
s_{\tetau}^2 = 10^{-3}$, $\Phi = \psun$.\\

\noindent{\bf Figure 8:} $\Psurv$ surface for $s_{2\temu}^2 = 5\times 10^{-3},  
s_{\tetau}^2 = 10^{-3}$, $\Phi = \psc$.\\

\noindent{\bf Figure 9:} $\Psurv$ surface for $s_{2\temu}^2 = 5\times 10^{-3},  
s_{\tetau}^2 = 0.4$, $\Phi = \psun$.\\

\noindent{\bf Figure 10:} $\Psurv$ surface for $s_{2\temu}^2 = 5\times 10^{-3},  
s_{\tetau}^2 = 0.4$, $\Phi = \psc$.\\

\noindent{\bf Figure 11:} Overhead projection of Figure~9 showing resonance boundary
and hierarchy breaking transition.\\

\noindent{\bf Figure 12:} Overhead projection of Figure~10 showing resonance boundary
and hierarchy breaking transition.\\

\pagebreak

\pagebreak

\pagebreak
 
\pagebreak
 
\pagebreak
 
\pagebreak
 
\pagebreak
 
\pagebreak
 
\pagebreak
 
\pagebreak
 
\pagebreak
 
\pagebreak
 
\end{document}